\documentclass[aps,prd,groupedaddress,twocolumn,amssymb,eqsecnum,showpacs,showkeys,epsfig,nofootinbib]{revtex4}
\usepackage{graphicx}
\usepackage{bm}
\usepackage{amsmath}
\usepackage[latin1]{inputenc}
\usepackage[english]{babel}
\usepackage{amssymb}
\usepackage{amsfonts}
\usepackage{epsfig}

\def \lleq {\lower0.9ex\hbox{ $\buildrel < \over \sim$} ~}
\def \ggeq {\lower0.9ex\hbox{ $\buildrel > \over \sim$} ~}

\def \beq  {\begin{equation}}
\def \eeq  {\end{equation}}
\def \ber  {\begin{eqnarray}}
\def \eer  {\end{eqnarray}}

\begin{document}
\newcommand{\newc}{\newcommand}
\newc{\ben}{\begin{eqnarray}}
\newc{\een}{\end{eqnarray}}
\newc{\be}{\begin{equation}}
\newc{\ee}{\end{equation}}
\newc{\ba}{\begin{eqnarray}}
\newc{\ea}{\end{eqnarray}}
\newc{\bea}{\begin{eqnarray*}}
\newc{\eea}{\end{eqnarray*}}
\newc{\D}{\partial}
\newc{\na}{\nabla}
\newc{\ie}{{\it i.e.} }
\newc{\eg}{{\it e.g.} }
\newc{\etc}{{\it etc.} }
\newc{\etal}{{\it et al.}}
\newcommand{\nn}{\nonumber}
\newc{\ra}{\rightarrow}
\newc{\lra}{\leftrightarrow}
\newc{\lsim}{\buildrel{<}\over{\sim}}
\newc{\gsim}{\buildrel{>}\over{\sim}}
\newc{\G}{\Gamma}
\title{Massive, massless and ghost modes of gravitational waves from higher-order gravity }

\author{Charalampos Bogdanos$^{1}$, Salvatore Capozziello$^{2,3}$, Mariafelicia De Laurentis$^{2,3}$, Savvas Nesseris$^{4}$}

\affiliation{\it $^1$ LPT, Universit$\acute{e}$ de Paris-Sud-11,
B$\hat{a}$t. 210, 91405 Orsay CEDEX, France \\  $^2$Dipartimento
di Scienze Fisiche, Università di Napoli {}`` Federico II'' and
$^3$INFN Sez. di Napoli, Compl. Univ. di Monte S. Angelo, Edificio G, Via Cinthia, I-80126, Napoli, Italy\\
$^4$The Niels Bohr International Academy, The Niels Bohr
Institute, Blegdamsvej 17, DK-2100, Copenhagen \O, Denmark}

\date{\today}
\begin{abstract}
We linearize the field equations for higher order theories that
contain scalar invariants other than the Ricci scalar. We find
that besides a massless spin-2 field (the standard graviton), the
theory contains also spin-0 and spin-2 massive modes with the
latter being, in general,  ghost modes. Then, we investigate the
possible detectability of such additional polarization modes of a
stochastic gravitational wave by ground-based and space
interferometric detectors. Finally, we extend the formalism of the
cross-correlation analysis, including the additional polarization
modes, and calculate the detectable energy density of the spectrum
for a stochastic background of the relic gravity waves that
corresponds to our model. For the situation considered here, we
find that these massive modes are certainly of interest for direct
detection by the LISA experiment.
\end{abstract}

\pacs{04.30, 04.30.Nk, 04.50.+h, 98.70.Vc}
\keywords{gravitational waves; alternative theories of gravity; cosmology}

\maketitle
\section{Introduction}
\label{uno}
Recently, the data analysis of interferometric gravitational wave
(GW) detectors has been started (for the current status of GWs
interferometers see \cite{acernese,willke,sigg,abbott,ando}) and
the scientific community aims at a first direct detection of GWs
in next years. The design and the construction of a number of
sensitive detectors for GWs is underway today. There are some
laser interferometers like the VIRGO detector, built in Cascina,
near Pisa, Italy, by a joint Italian-French collaboration, the GEO
600 detector built in Hannover, Germany, by a joint Anglo-German
collaboration, the two LIGO detectors built in the United States
(one in Hanford, Washington and the other in Livingston,
Louisiana) by a joint Caltech-MIT collaboration, and the TAMA 300
detector, in Tokyo, Japan.

Many  detectors are currently in operation too, and several
interferometers  are in a phase of planning and proposal stages
(for the current status of gravitational waves experiments see
\cite{ligo,virgo,lisa}). The results of these detectors will have
a fundamental impact on astrophysics and gravitational physics and
will be important for a better knowledge of the Universe and
either to confirm or rule out the physical consistency of General
Relativity or any other theory of gravitation \cite{will}. Several
issues coming from Cosmology and Quantum Field Theory suggest to
extend the Einstein General Relativity (GR), in order to cure
several shortcomings emerging from astrophysical observations and
fundamental physics. For example, problems in early time cosmology
led to the conclusion that the Standard Cosmological Model could
be inadequate to describe the Universe at extreme regimes. In
fact, GR does not work at the fundamental level, when one wants to
achieve a full quantum description of space-time (and then of
gravity).

Given these facts and the lack of a final self-consistent Quantum
Gravity Theory, alternative theories of gravity have been pursued
as part of a semi-classical scheme where GR and its positive
results should be recovered. The approach of Extended Theories of
Gravity (ETGs) based on corrections and enlargements of the
Einstein scheme, have become a sort of paradigm in the study of
the gravitational interaction. Beside fundamental physics
motivations, these theories have received a lot of interest in
cosmology since they ``naturally" exhibit inflationary behavior
which can overcome the shortcomings of standard cosmology. The
related cosmological models seem realistic and capable of coping
with observations. ETGs are starting to play an interesting role
to describe today's observed Universe. In fact, the good quality
data of last decade has made it possible to shed new light on the
effective picture of the Universe.

From an astrophysical point of view, ETGs do not require finding
candidates for dark energy and dark matter at the fundamental
level; the approach starts from taking into account only the
``observed" ingredients (i.e. gravity, radiation and baryonic
matter); it is in full agreement with the early spirit of a GR
that could not act in the same way at all scales. For example, it
is possible to show that several scalar-tensor and $f(R)$-models
(where $f$ is a generic function of the Ricci scalar $R$) agree
with observed cosmology, extragalactic and galactic observations
and Solar System tests, and give rise to new effects capable of
explaining the observed acceleration of the cosmic fluid and the
missing matter effect of self-gravitating structures without
considering dark energy and dark matter. For comprehensive reviews
on the argument, see \cite{CF}.

At a fundamental level,  detecting new gravitational modes could
be a sort of {\it experimentum crucis} in order to discriminate
among theories since this fact would be the ``signature" that GR
should be enlarged or modified \cite{bellucci,elizalde}.

The outline of the paper is as follows. In  Sect. \ref{due},  the
general action of the class of theories under consideration is
introduced. Then we will linearize them around a Minkowski
background  to find the  modes of the metric perturbations. In
Sect. \ref{tre}, we take into account  the various polarizations
of the massless and massive modes, while in Sect. \ref{quattro} we
investigate the response of  a single detector to a GW propagating
in  certain direction with each polarization mode. In Sect.
\ref{cinque}, we discuss the spectrum of the  GW stochastic
background where also further modes are considered. Conclusions
are drawn in Sect. \ref{sei}.

\section{Higher order gravity}
\label{due}
Let us generalize the action of GR by adding curvature invariants
other than the Ricci scalar. Specifically, we will consider the
action \footnote{Conventions: $g_{ab}=(-1,1,1,1),~~
R^a_{bcd}=\G^a_{bd,c}-\G^a_{bc,d}+...~,~~ R_{ab}=R^c_{acb},~~
G_{ab}=8 \pi G_N T_{ab}$ and all indices run from 0 to 3.}

 \be S=\int
d^4x\sqrt{-g} f(R,P,Q) \ee where \ba && P\equiv R_{ab}R^{ab}\nn \\
&&Q\equiv R_{abcd}R^{abcd} \ea

Varying with respect to the metric one gets the field equations
\cite{Carroll:2004de}:

 \ba FG_{\mu\nu}&=&\frac{1}{2}g_{\mu\nu}\left(f-
R~F\right)-(g_{\mu\nu}\Box-\na_\mu\na_\nu)F\nn\\&&
-2\left(f_P R^a_\mu
R_{a\nu}+f_Q~R_{abc\mu}R^{abc}_{~~~\nu} \right)\nn\\&&-
g_{\mu\nu}\na_a\na_b(f_P R^{ab})-\Box (f_P
R_{\mu\nu})\nn\\ &&+2\na_a\na_b\left(f_P~R^a_{~(\mu}\delta^b_{~\nu)}+2
f_Q~R^{a~~~~b}_{~(\mu\nu)}\right)\nn\\\label{fieldeqs}\ea

where we have set
  \be F\equiv\frac{\D f}{\D R}, ~~~f_P\equiv\frac{\D f}{\D P}, ~~~f_Q\equiv\frac{\D
f}{\D Q} \ee and $\Box=g^{ab}\na_a\na_b$ is the d'Alembert
operator while the notation $T_{(ij)}=\frac{1}{2}(T_{ij}+T_{ji})$
denotes symmetrization with respect to the indices $(i,j)$.

Taking the trace of eq. (\ref{fieldeqs}) we find:
 \ba && \Box\left(F+\frac{f_P}{3}
R\right)=\nonumber\\&&
\frac{1}{3}\left( 2 f-RF-2 \na_a\na_b((f_P+2f_Q)R^{ab})-2
(f_P P+f_Q Q)\right)\nn\\ \label{trace}\ea

Expanding the third term on the RHS of (\ref{trace}) and using the
purely geometrical identity $G^{ab}_{~~;b}=0$ we get: \ba
&&\Box\left(F+\frac{2}{3}(f_P+f_Q) R\right)=
\frac{1}{3}\times\nn\\ && [2
f-RF-2R^{ab}\na_a\na_b(f_P+2f_Q)-R\Box(f_P+2f_Q)\nn\\ &&-2 (f_P
P+f_Q Q)] \label{trace1}\ea If we define \ba \Phi & \equiv&
F+\frac{2}{3}(f_P+f_Q) R \label{phidef} \\ && \textrm{and} ~~~ \nn
\\\frac{dV}{d\Phi} & \equiv& \textrm{RHS ~of~ (\ref{trace1})}\nn \ea
then we get a Klein-Gordon equation for the scalar field $\Phi$:
\be \Box \Phi = \frac{dV}{d\Phi} \ee In order to find the various
modes of the gravity waves of this theory we need to linearize
gravity around a Minkowski background:
\ba g_{\mu\nu}&=&\eta_{\mu\nu}+h_{\mu\nu} \nn \\
\Phi&=&\Phi_0+\delta \Phi \ea Then from eq. (\ref{phidef}) we get

\be \delta \Phi=\delta F+\frac{2}{3}(\delta f_P+\delta f_Q) R_0+
\frac{2}{3}(f_{P0}+f_{Q0}) \delta R \label{pertphi1}\ee where $R_0
\equiv R(\eta_{\mu\nu})=0$ and similarly $f_{P0}=\frac{\D f}{\D
P}|_{\eta_{\mu\nu}}$ (note that the 0 indicates evaluation with
the Minkowski metric) which is either constant or zero. By $\delta
R$ we denote the first order perturbation on the Ricci scalar
which, along with the perturbed parts of the Riemann and Ricci
tensors, are given by (see for example Ref.\cite{Carroll:1997ar}):

\ba \delta R_{\mu\nu\rho\sigma}&=&\frac{1}{2}\left(\D_\rho \D_\nu
h_{\mu \sigma}+\D_\sigma \D_\mu h_{\nu \rho}-\D_\sigma \D_\nu
h_{\mu
\rho}-\D_\rho \D_\mu h_{\nu \sigma} \right) \nn\\
\delta R_{\mu\nu} &=& \frac{1}{2}\left(\D_\sigma \D_\nu
h^\sigma_{~\mu}+\D_\sigma \D_\mu h^\sigma_{~\nu}-\D_\mu \D_\nu
h-\Box h_{\mu \nu} \right)\nn\\
\delta R &=& \D_\mu \D_\nu h^{\mu \nu}-\Box h\nn\ea where
$h=\eta^{\mu \nu} h_{\mu \nu}$. The first term of eq.
(\ref{pertphi1}) is \be \delta F=\frac{\D F}{\D R}|_0~\delta
R+\frac{\D F}{\D P}|_0~\delta P+\frac{\D F}{\D Q}|_0~\delta Q \ee
However, since $\delta P$ and $\delta Q$ are second order we get
$\delta F\simeq F_{,R0}~ \delta R $ and \be \delta \Phi
=\left(F_{,R0} +\frac{2}{3} (f_{P0}+f_{Q0})\right) \delta R
\label{pertphi2}\ee Finally, from eq. (\ref{trace1}) we get the
Klein-Gordon equation for the scalar perturbation $\delta \Phi$

\ba \Box \delta \Phi&=&\frac{1}{3}\frac{F_0}{F_{,R0} +\frac{2}{3}
(f_{P0}+f_{Q0})}\delta \Phi-\nn\\ &&\frac{2}{3}\delta
{R}^{ab}\D_a\D_b(f_{P0}+2f_{Q0})-\frac{1}{3}\delta
{R}\Box(f_{P0}+2f_{Q0})\nn \\ &=& m_s^2 \delta \Phi
\label{kgordon1}\nn\\ \ea The last two terms in the first line are
actually are zero since the terms $f_{P0}$, $f_{Q0}$ are constants
and we have defined the scalar mass as $m_s^2\equiv
\frac{1}{3}\frac{F_0}{F_{,R0} +\frac{2}{3} (f_{P0}+f_{Q0})}$.

Perturbing the field equations (\ref{fieldeqs}) we get: \ba &&
F_0(\delta{R}_{\mu\nu}-\frac{1}{2}\eta_{\mu\nu}
\delta{R})=\nn\\ &&-(\eta_{\mu\nu}\Box -\D_\mu\D_\nu)(\delta
\Phi-\frac{2}{3}(f_{P0}+f_{Q0})\delta{R})\nn
\\&&-\eta_{\mu\nu} \D_a\D_b (f_{P0} \delta{R}^{ab})-\Box(f_{P0}
\delta{R}_{\mu\nu})\nn\\ &&+2
\D_a\D_b(f_{P0}~\delta{R}^a_{~(\mu}\delta^b_{~\nu)}+2
f_{Q0}~\delta{R}^{a~~~~b}_{~(\mu\nu)})\nn\\ \ea It is convenient to
work in Fourier space so that for example $\D_\gamma
h_{\mu\nu}\rightarrow i k_\gamma h_{\mu\nu}$ and $\Box h_{\mu\nu}
\rightarrow -k^2 h_{\mu\nu}$. Then the above equation becomes \ba
&& F_0(\delta{R}_{\mu\nu}-\frac{1}{2}\eta_{\mu\nu}
\delta{R})=\nn\\ &&(\eta_{\mu\nu}k^2 -k_\mu k_\nu)(\delta
\Phi-\frac{2}{3}(f_{P0}+f_{Q0})\delta{R})\nn
\\&&+\eta_{\mu\nu} k_a k_b (f_{P0} \delta{R}^{ab})+k^2(f_{P0}
\delta{R}_{\mu\nu})\nn\\ &&-2 k_a
k_b(f_{P0}~\delta{R}^a_{~(\mu}\delta^b_{~\nu)})-4 k_a k_b(
f_{Q0}~\delta{R}^{a~~~~b}_{~(\mu\nu)})\nn\\ \label{fields2}\ea We
can rewrite the metric perturbation as \be
h_{\mu\nu}=\bar{h}_{\mu\nu}-\frac{\bar{h}}{2}~
\eta_{\mu\nu}+\eta_{\mu\nu} h_f \label{gauge}\ee and use our gauge
freedom to define to demand that the usual conditions hold $\D_\mu
\bar{h}^{\mu\nu} =0$ and $\bar{h}=0$. The first of these
conditions implies that $k_\mu \bar{h}^{\mu\nu} =0$ while the
second that \ba h_{\mu\nu}&=&\bar{h}_{\mu\nu}+\eta_{\mu\nu} h_f \nn \\
h&=&4 h_f\ea With these in mind we have:
 \ba \delta
R_{\mu\nu}&=&\frac{1}{2}\left(2k_\mu k_\nu h_f+k^2 \eta _{\mu\nu}
h_f+k^2 \bar{h}_{\mu\nu}\right) \nn\\
\delta R &=& 3k^2 h_f\nn\\
k_\alpha k_\beta ~\delta
R^{\alpha~~~~~\beta}_{~~(\mu\nu)~}&=&-\frac{1}{2}\left((k^4
\eta_{\mu\nu}-k^2 k_\mu k_\nu)h_f+k^4 \bar{h}_{\mu\nu}\right)\nn\\
k_a k_b~\delta{R}^a_{~(\mu}\delta^b_{~\nu)}&=&\frac{3}{2}k^2k_\mu
k_\nu h_f \nn\\ \label{results1}\ea Using equations
(\ref{gauge})-(\ref{results1}) into (\ref{fields2}) and after some
algebra we get: \ba
&&\frac{1}{2}\left(k^2-k^4\frac{f_{P0}+4f_{Q0}}{F_0}\right)\bar{h}_{\mu\nu}=\nn\\
&&(\eta_{\mu\nu}k^2 -k_\mu k_\nu)\frac{\delta \Phi}{F_0}
+(\eta_{\mu\nu}k^2 -k_\mu k_\nu)h_f \nn\\\ea Defining $h_f\equiv
-\frac{\delta \Phi}{F_0}$ we find the equation for the
perturbations: \be
\left(k^2+\frac{k^4}{m^2_{spin2}}\right)\bar{h}_{\mu\nu}=0
\label{solution} \ee where we have defined $m^2_{spin2}\equiv
-\frac{F_0}{f_{P0}+4f_{Q0}}$, while from eq. (\ref{kgordon1}) we
get: \be \Box h_f=m_s^2 h_f \label{kgordon3}\ee From equation
(\ref{solution}) it is easy to see that we have a modified
dispersion relation which corresponds to a massless spin-2 field
($k^2=0$) and a massive spin-2 ghost mode
$k^2=\frac{F_0}{\frac{1}{2}f_{P0}+2f_{Q0}}\equiv -m^2_{spin2}$
with mass $m^2_{spin2}$. To see this, note that the propagator for
$\bar{h}_{\mu\nu}$ can be rewritten as \be G(k) \propto
\frac{1}{k^2}-\frac{1}{k^2+m^2_{spin2}} \ee Clearly the second
term has the opposite sign, which indicates the presence of a
ghost, and this agrees with the results found in the literature
for this class of theories
\cite{Nunez:2004ts,Chiba:2005nz,Stelle:1977ry}.

Also, as a sanity check, we can see that for the Gauss-Bonnet term
$\mathcal{L}_{GB}=Q-4P+R^2$ we have $f_{P0}=-4$ and $f_{Q0}=1$.
Then, equation (\ref{solution}) simplifies to $k^2
\bar{h}_{\mu\nu}=0$ and in this case we have no ghosts as
expected.

The solution to eqs. (\ref{solution}) and (\ref{kgordon3}) can be
written in terms of plane waves \ba \bar{h}_{\mu\nu}&=&A_{\mu\nu}
(\overrightarrow{p}) \cdot  exp(ik^\alpha x_\alpha)+cc \label{pw1}
\ea \ba h_f &=& a(\overrightarrow{p}) \cdot exp(iq^\alpha
x_\alpha)+cc\label{pw2} \ea where

\begin{equation}
\begin{array}{ccc}
k^{\alpha}\equiv(\omega_{m_{spin2}},\overrightarrow{p}) &  & \omega_{m_{spin2}}=\sqrt{m_{spin2}^{2}+p^{2}}\\
\\q^{\alpha}\equiv(\omega_{m_s},\overrightarrow{p}) &  & \omega_{m_s}=\sqrt{m_s^{2}+p^{2}}.\end{array}\label{eq: k e q}\end{equation} and
where $m_{spin2}$ is zero (non-zero) in the case of massless
(massive) spin-2 mode and the polarization tensors $A_{\mu\nu}
(\overrightarrow{p})$ can be found in Ref. \cite{vanDam:1970vg}
(see equations (21)-(23)). In eqs. (\ref{solution}) and
(\ref{pw1}) the equation and the solution for the standard waves
of General Relativity \cite{gravitation} have been obtained, while
eqs. (\ref{kgordon3}) and (\ref{pw2}) are respectively the
equation and the solution for the massive mode (see also
\cite{felix}).

The fact that the dispersion law for the modes of the massive
field $h_{f}$ is not linear has to be emphasized. The velocity of
every {}``ordinary'' (i.e. which arises from General Relativity)
mode $\bar{h}_{\mu\nu}$ is the light speed $c$, but the dispersion
law (the second of eq. (\ref{eq: k e q})) for the modes of $h_{f}$
is that of a massive field which can be discussed like a
wave-packet \cite{felix}. Also, the group-velocity of a
wave-packet of $h_{f}$ centered in $\overrightarrow{p}$ is

\begin{equation}
\overrightarrow{v_{G}}=\frac{\overrightarrow{p}}{\omega},\label{eq: velocita' di gruppo}\end{equation}

which is exactly the velocity of a massive particle with mass $m$
and momentum $\overrightarrow{p}$.

From the second of eqs. (\ref{eq: k e q}) and eq. (\ref{eq: velocita' di gruppo})
it is simple to obtain:

\begin{equation}
v_{G}=\frac{\sqrt{\omega^{2}-m^{2}}}{\omega}.\label{eq: velocita' di gruppo 2}\end{equation}

Then, wanting a constant speed of the wave-packet, it has to be
\cite{felix}

\begin{equation}
m=\sqrt{(1-v_{G}^{2})}\omega.\label{eq: relazione massa-frequenza}\end{equation}

Now, before we proceed with the analysis, we should discuss the
phenomenological limitations to the mass of the GW \cite{corda1}.
Taking into account the fact that the GW needs a frequency which
falls in the range for both of space based and earth based
gravitational antennas, that is the interval $10^{-4}Hz\leq
f\leq10KHz$
\cite{acernese,willke,sigg,abbott,ando,tatsumi,lisa1,lisa2}, a
quite strong limitation will arise. For a massive GW, from
\cite{capozzcorda} it is:

\begin{equation}
2\pi f=\omega=\sqrt{m^{2}+p^{2}},\label{eq: frequenza-massa}\end{equation}

were $p$ is the momentum. Thus, it needs

\begin{equation}
0eV\leq m\leq10^{-11}eV.\label{eq: range di massa}\end{equation}

A stronger limitation is given by requirements of cosmology and
Solar System tests on extended theories of gravity. In this case
it is

\begin{equation}
0eV\leq m\leq10^{-33}eV.\label{eq: range di massa 2}\end{equation}

For these light scalars, their effect can be still discussed as a
coherent GW.

\section{Polarization states of gravitational waves}
\label{tre}

Considering the above equations, we can note that there are two
conditions for eq. (\ref{kgordon1}) that depend on the value of
$k^2$. In fact  we can have a $k^2=0$ mode that corresponds to a
massless spin-2 field with two independent polarizations plus a
scalar mode, while if we have $k^2\neq0$ we have a massive spin-2
ghost mode and there are five independent polarization tensors
plus a scalar mode. First, lets consider the case where the spin-2
field is massless.

Taking $\overrightarrow{p}$ in the $z$ direction, a gauge in which
only $A_{11}$, $A_{22}$, and $A_{12}=A_{21}$ are different to zero
can be chosen. The condition $\bar{h}=0$ gives $A_{11}=-A_{22}$.
 In this frame we may take the bases of  polarizations defined in this way\footnote{The polarizations are
defined in our 3-space, not in a spacetime with extra
dimensions. Each polarization mode is orthogonal to one
another and is normalized $e_{\mu\nu}e^{\mu\nu} =2\delta$. Note that other modes are not traceless, in contrast to the ordinary
plus and cross polarization modes in GR.}
\begin {equation}
e_{\mu\nu}^{(+)}=\frac{1}{\sqrt{2}}\left(
\begin{array}{ccc}
1 & 0 & 0 \\
0 & -1 & 0 \\
0 & 0 & 0
\end{array}
\right),\nonumber\qquad e_{\mu\nu}^{(\times)}=\frac{1}{\sqrt{2}}\left(
\begin{array}{ccc}
0 & 1 & 0 \\
1 & 0 & 0 \\
0 & 0 & 0
\end{array}
\right)\nonumber
\end{equation}
\begin {equation}
e_{\mu\nu}^{(s)}=\frac{1}{\sqrt{2}}\left(
\begin{array}{ccc}
0 & 0 & 0 \\
0 & 0 & 0 \\
0 & 0 & 1
\end{array}\right)
\end{equation}


Now, putting these equations in eq. (\ref{gauge}), it results

\ba h_{\mu\nu}(t,z)&=&A^{+}(t-z)e_{\mu\nu}^{(+)}
+A^{\times}(t-z)e_{\mu\nu}^{(\times)}\nn
\\&+&h_{s}(t-v_{G}z)e_{\mu\nu}^{s}\label{eq: perturbazione
totale}\ea

The terms
$A^{+}(t-z)e_{\mu\nu}^{(+)}+A^{\times}(t-z)e_{\mu\nu}^{(\times)}$
describe the two standard polarizations of gravitational waves
which arise from General Relativity, while the term
$h_{s}(t-v_{G}z)\eta_{\mu\nu}$ is the massive field arising from
the generic high order $f(R$) theory.

When the spin-2 field is massive,
 we have that the bases of the six polarizations are defined by
\begin {equation}
e_{\mu\nu}^{(+)}=\frac{1}{\sqrt{2}}\left(
\begin{array}{ccc}
1 & 0 & 0 \\
0 & -1 & 0 \\
0 & 0 & 0
\end{array}
\right),\nonumber\qquad e_{\mu\nu}^{(\times)}=\frac{1}{\sqrt{2}}\left(
\begin{array}{ccc}
0 & 1 & 0 \\
1 & 0 & 0 \\
0 & 0 & 0
\end{array}
\right)\nonumber
\end{equation}

\begin {equation}
e_{\mu\nu}^{(B)}=\frac{1}{\sqrt{2}}\left(
\begin{array}{ccc}
0 & 0 & 1 \\
0 & 0 & 0 \\
1 & 0 & 0
\end{array}
\right),\nonumber\qquad e_{\mu\nu}^{(C)}=\frac{1}{\sqrt{2}}\left(
\begin{array}{ccc}
0 & 0 & 0 \\
 0 & 0 & 1 \\
0 & 1 & 0
\end{array}
\right)\nonumber
\end{equation}
\begin {equation}
e_{\mu\nu}^{(D)}=\frac{\sqrt{2}}{3}\left(
\begin{array}{ccc}
\frac{1}{2} & 0 & 0 \\
0 & \frac{1}{2} & 0 \\
0 & 0 & -1
\end{array}
\right),\nonumber\qquad e_{\mu\nu}^{(s)}=\frac{1}{\sqrt{2}}\left(
\begin{array}{ccc}
0 & 0 & 0 \\
0 & 0 & 0 \\
0 & 0 & 1
\end{array}
\right)\label{tensorpol}\nonumber
\end{equation}

and the amplitude can be written in
terms of the 6 polarization states as

\ba
&&h_{\mu\nu}(t,z)=A^{+}(t-v_{G_{s2}} z)e_{\mu\nu}^{(+)}+A^{\times}(t-v_{G_{s2}} z)e_{\mu\nu}^{(\times)}\nn\\
&&+B^{B}(t-v_{G_{s2}} z)e_{\mu\nu}^{(B)}+C^{C}(t-v_{G_{s2}} z)e_{\mu\nu}^{(C)}\nn\\
&&+D^{D}(t-v_{G_{s2}}
z)e_{\mu\nu}^{(D)}+h_{s}(t-v_{G}z)e_{\mu\nu}^{s}.\nn\\\ea where
$v_{G_{s2}}$ is the group velocity of the massive spin-2 field and
is given by \be
v_{G_{s2}}=\frac{\sqrt{\omega^{2}-m_{s2}^{2}}}{\omega}.\label{spin2group}\ee

The first two polarizations are the same as in the massless case,
inducing tidal deformations on the x-y plane. In Fig.1, we
illustrate how each GW polarization affects test masses arranged
on a circle.
\begin{figure}
\begin{center}
\leavevmode
\centerline{\epsfig{figure=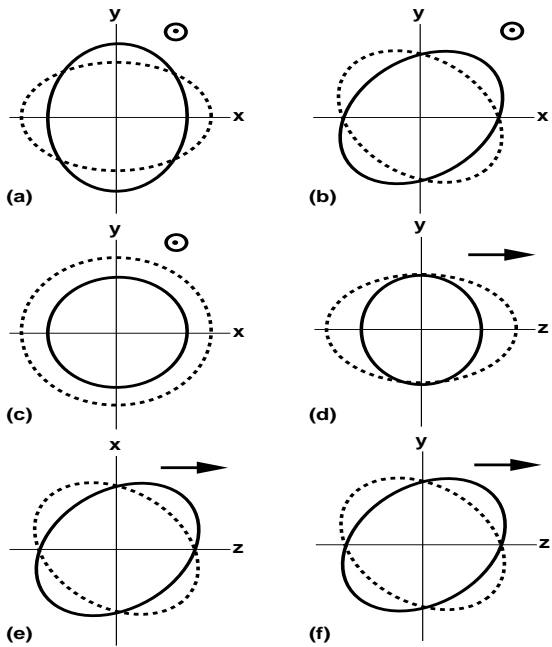,angle=360,height=8.5cm,width=7.25cm}}
\caption{The six polarization modes of  gravitational waves. The picture shows
the displacement that each mode induces on a
sphere of test particles at the moments of different phases by
$\pi$. The wave propagates out of the plane in (a), (b), (c), and
it propagates in the plane in (d), (e) and (f). Where in (a) and
(b) we have respectively the plus mode and cross mode, in (c) the
scalar mode, in (d), (e) and (f) the D, B and C mode.}
\end{center}
\label{fig1}
\end{figure}

The presence of the ghost mode may seem as a pathology of the
theory from a purely quantum-mechanical approach. There are
several reasons to consider such a mode as problematic if we wish
to pursuit the particle picture interpretation of the metric
perturbations. The ghost mode can be viewed as either a particle
state of positive energy and negative probability density, or a
positive probability density state with a negative energy. In the first
case, allowing the presence of such a particle will quickly induce
violation of unitarity. The negative energy scenario leads to a
theory where there is no minimum energy and the system thus
becomes unstable. The vacuum can decay into pairs of ordinary and
ghost gravitons leading to a catastrophic instability.

One way out of such problems is to impose a very weak coupling of
the ghost with the rest of the particles in the theory, such that
the decay rate of the vacuum will become comparable to the inverse
of the Hubble scale. The present vacuum state will then appear to
be sufficiently stable. This is not a viable option in our theory,
since the ghost state comes in the gravitational sector, which is
bound to couple to all kinds of matter present and it seems
physically and mathematically unlikely for the ghost graviton to
couple differently than the ordinary massless graviton does.
Another option is to assume that this picture does not hold up to
arbitrarily high energies and that at some cutoff scale
$M_{cutoff}$ the theory gets modified appropriately as to ensure a
ghost-free behavior and a stable ground state. This can happen for
example if we assume that Lorentz invariance is violated at
$M_{cutoff}$, thereby restricting any potentially harmful decay
rates \cite{Emparan:2005gg}.

However, there is no guaranty that theories of modified gravity
such as the one investigated here are supposed to hold up to
arbitrary energies. Such models are plagued at the quantum level
by the same problems as ordinary General Relativity, i.e. they are
non-renormalizable. It is therefore not necessary for them to be
considered as genuine candidates for a quantum gravity theory and
the corresponding ghost particle interpretation becomes rather
ambiguous. At the purely classical level, the perturbation $h_{\mu
\nu}$ should be viewed as nothing more than a tensor representing
the ``stretching'' of spacetime away from flatness. A ghost mode
then makes sense as just another way of propagating this
perturbation of the spacetime geometry, one which carries the
opposite sign in the propagator than an ordinary massive graviton
would.

Viewed in this way, the presence of the massive ghost graviton
will induce on an interferometer the same effects as an ordinary
massive graviton transmitting the perturbation, but with the
opposite sign in the displacement. Tidal stretching from a
polarized wave on the polarization plane will be turned into
shrinking and vice-versa. This signal will in the end be a
superposition of the displacements coming from the ordinary
massless spin-2 graviton and the massive ghost. Since these induce
two competing effects, this will lead to a less pronounced signal
than the one we would expect if the ghost mode was absent, setting
in this way less severe constraints on the theory. However, the presence of the new
modes will also affect the total energy density carried by the gravitational
waves and this may also appear as a candidate signal in stochastic backgrounds, as we will see in the following.

\section{Gravitational waves propagating in a certain direction and the possible detector response }
\label{quattro}
\begin{figure}[h]
\begin{center}
\includegraphics[width=6.5cm]{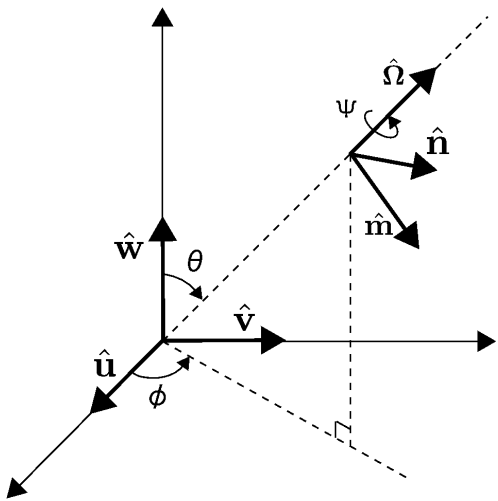}
\caption{The  coordinate systems used to calculate the polarization tensors and the pictorial view of the coordinate transformation.}
\label{fig2}
\end{center}
\end{figure}

Let us consider now now
the possible response of a  detector revealing  GWs coming from a certain
direction.  It is important to stress that the detector output
 depends on the GW amplitude that is determined by a specific
theoretical model. However, one can study the detector response to each GW
polarization without specifying, a priori,  the theoretical model. Following \cite{abio,bonasia,babusci,vicere,leaci,Maggiore} the
angular pattern function of a detector to GWs is given by
\begin{eqnarray}
F_A (\hat{\mathbf{\Omega}}) &=& \mathbf{D} : \mathbf{e}_A
(\hat{\mathbf{\Omega}})\:,
\label{eq2} \\
\mathbf{D} &=&  \frac{1}{2}\left[ \hat{\mathbf{u}} \otimes \hat
{\mathbf{u}}- \hat{\mathbf{v}}
\otimes \hat{\mathbf{v}}\right]\:,
\nonumber
\end{eqnarray}
here $A=+,\times,B,C,D,s$.  The symbol :  is
contraction between tensors.  $\mathbf{D}$ is the
{\it detector tensor} representing  the response of a
laser-interferometric detector. It maps the  metric
perturbation in a  signal on the detector. The vectors
$\hat{\mathbf{u}}$ and $\hat{\mathbf{v}}$ are unitary  and orthogonal to each
other.  They are directed to each detector arm and form an
orthonormal coordinate system with the unit vector
$\hat{\mathbf{w}}$ (see  Fig.\,\ref{fig2}).
$\hat{\mathbf{\Omega}}$ is the  vector directed along the GW
propagation.
Eq.\,(\ref{eq2}) holds only when the arm length of the
detector is  smaller and smaller than the GW wavelength  that we are taking into account.
 This is relevant for  dealing with
ground-based laser interferometers but this condition could not be valid when dealing with space interferometers like LISA.

A standard orthonormal coordinate system for the detector is
\begin{equation}
\left\{
\begin{array}{lll}
\displaystyle
\hat{\mathbf{u}} = (1,0,0)
\\
\displaystyle
\hat{\mathbf{v}} = (0,1,0)
\\
\displaystyle
\hat{\mathbf{w}} = (0,0,1)
\end{array}
\right. \;. \nonumber
\end{equation}
On the other hand,  the coordinate system for the GW, rotated by angles $(\theta, \phi)$,  is given by
\begin{equation}
\left\{
\begin{array}{lll}
\displaystyle
\hat{\mathbf{u}}^{\prime} =(\cos \theta \cos \phi , \cos \theta \sin \phi , -\sin \theta)
\\
\displaystyle
\hat{\mathbf{v}}^{\prime} = (- \sin \phi , \cos \phi , 0)
\\
\displaystyle
\hat{\mathbf{w}}^{\prime} = (\sin \theta \cos \phi , \sin \theta \sin \phi , \cos \theta)
\end{array}
\right. \;. \nonumber
\end{equation}
The
rotation with respect to the angle $\psi$, around the
GW-propagating axis, gives the most general choice for the coordinate system, that is
\begin{equation}
\left\{
\begin{array}{lll}
\displaystyle
\hat{\mathbf{m}} = \hat{\mathbf{u}}^{ \prime} \cos \psi + \hat{\mathbf{v}}^{\prime} \sin \psi\\
\displaystyle
\hat{\mathbf{n}} = - \hat{\mathbf{v}}^{ \prime} \sin \psi + \hat{\mathbf{u}} ^{\prime} \cos \psi
\\
\displaystyle
\hat{\mathbf{\Omega}} = \hat{\mathbf{w}}^{ \prime}
\end{array}
\right. \;.
\nonumber
\end{equation}
Coordinates
$(\hat{\mathbf{u}},\hat{\mathbf{v}},\hat{\mathbf{w}})$ are related
to the coordinates
$(\hat{\mathbf{m}},\hat{\mathbf{n}},\hat{\mathbf{\Omega}})$ by the
rotation angles ($\phi,\,\theta,\,\psi$), as in
Fig.\,\ref{fig2}. By thevectors $\hat{\mathbf{m}}$,
$\hat{\mathbf{n}}$, and $\hat{\mathbf{\Omega}}$, the polarization
tensors are
\begin{eqnarray}
\mathbf{e}_{+} &=& \frac{1}{\sqrt{2}}\left(\hat{\mathbf{m}} \otimes \hat{\mathbf{m}} -\hat{\mathbf{n}} \otimes \hat{\mathbf{n}}\right) \;, \nonumber \\
\mathbf{e}_{\times} &=& \frac{1}{\sqrt{2}}\left( \hat{\mathbf{m}} \otimes \hat{\mathbf{n}} +\hat{\mathbf{n}} \otimes \hat{\mathbf{m}}\right) \;, \nonumber \\
\mathbf{e}_{B} &=& \frac{1}{\sqrt{2}}\left(\hat{\mathbf{m}} \otimes \hat{\mathbf{\Omega}} +\hat{\mathbf{\Omega}} \otimes \hat{\mathbf{m}}\right) \;, \nonumber \\
\mathbf{e}_{C} &=&\frac{1}{\sqrt{2}} \left(\hat{\mathbf{n}} \otimes \hat{\mathbf{\Omega}} +\hat{\mathbf{\Omega}} \otimes \hat{\mathbf{n}}\right) \;. \nonumber\\
\mathbf{e}_{D} &=&\frac{\sqrt{3}}{2}\left( \hat{\mathbf{\frac{m}{2}}} \otimes \hat{\mathbf{\frac{m}{2}}} + \hat{\mathbf{\frac{n}{2}}} \otimes \hat{\mathbf{\frac{n}{2}}}+ \hat{\mathbf{\Omega}} \otimes \hat{\mathbf{\Omega}} \right) \;,  \nonumber \\
\mathbf{e}_{s} &=& \frac{1}{\sqrt{2}}\left( \hat{\mathbf{\Omega}} \otimes \hat{\mathbf{\Omega}}\right) \;,  \nonumber \end{eqnarray}

Taking into account  Eqs.\,(\ref{eq2}), the angular patterns for
each polarization are
\begin{eqnarray}
F_{+}(\theta, \phi, \psi) &=&  \frac{1}{\sqrt{2}} (1+ \cos ^2 \theta ) \cos 2\phi \cos 2 \psi \nonumber \\
&& - \cos \theta \sin 2\phi \sin 2 \psi \;, \nonumber\\
F_{\times}(\theta, \phi, \psi) &=& - \frac{1}{\sqrt{2}} (1+ \cos ^2 \theta ) \cos 2\phi \sin 2 \psi \nonumber \\
&& -  \cos \theta \sin 2\phi \cos 2 \psi \;, \nonumber\\
F_{B}(\theta, \phi, \psi) &=& \sin \theta \,(\cos \theta \cos 2 \phi \cos \psi -\sin 2\phi \sin \psi) \;, \nonumber \\
F_{C}(\theta, \phi, \psi) &=& \sin \theta \,(\cos \theta \cos 2 \phi \sin \psi +\sin 2\phi \cos \psi) \;, \nonumber \\
F_{D}(\theta, \phi) &=& \frac{ \sqrt{3}}{32} \cos 2 \phi \left(6 \sin ^2\theta +(\cos 2 \theta +3) \cos 2 \psi \right)\;, \nonumber \\
F_{s}(\theta, \phi) &=& \frac{1}{\sqrt{2}} \sin^2 \theta \cos 2\phi \;. \nonumber
\label{eq5}
\end{eqnarray}

\begin{figure}[t]
\centering
{\vspace{0cm}\includegraphics[width=0.48\textwidth]{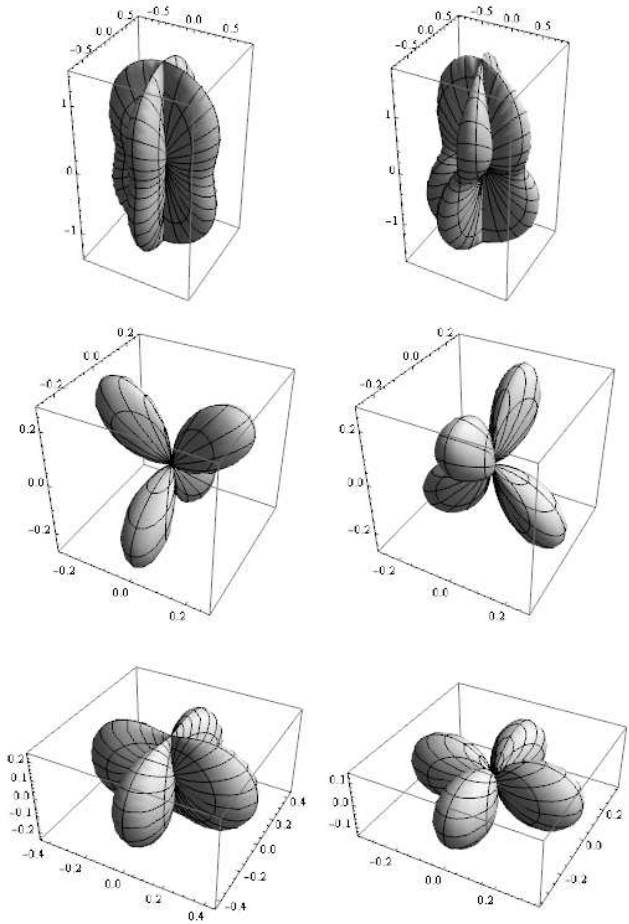}}
\caption{Plots along the panel lines from left to right of angular
pattern functions of a detector for each polarization. From left
plus mode $F_+$, cross mode $F_{\times}$,  B mode $F_{B}$,  C mode
$F_{C}$, D mode $F_{D}$,and scalar mode $F_{s}$. The angular
pattern function of the $F_B$ and $F_C$ mode is the same except
for a rotation. } \label{fig:3}
\end{figure}

The angular pattern functions for each  polarization are plotted in
Fig.\,\ref{fig:3}. These results, also if we have considered a different model,  are consistent, for example,  with those
 in \cite{abio,nishi,tobar}. Another step is now to consider the stochastic background of GWs in order to test the possible detectability of such further contributions in gravitational radiation.

\section{The stochastic background of gravitational waves}
\label{cinque}
The contributions to the gravitational radiation coming from higher order gravity could be efficiently selected if it  would be possible to investigate gravitational sources in extremely strong field regimes. In such a case, the further polarizations coming from the higher order contributions  could be, in principle, investigated by the response of a single GW detector described above. However, this situation seems extremly  futuristic at the moment so the only realistic approach to investigate these  further contribution seems the cosmological background, in particular, the stochastic background of GWs.
Such a GW background can be roughly
divided into two classes of phenomena: the  background generated by
 the incoherent superposition of gravitational radiation
emitted by large populations of astrophysical sources (hard to
be resolved individually~\cite{FP}), and the primordial GW
background generated by processes  in the early cosmological eras \cite{alebon}.
Primordial components of such background are
interesting, since they  carry  information on
the primordial Universe and, on the other hand, can give information on the gravitational interaction at that epochs \cite{francaviglia,cordafelix}.
The physical process of GW production has been analyzed, for example, in
\cite{Allen,Grishchuk,Allen2} but only for the first two standard tensorial
components of Eq. (\ref{eq: perturbazione totale}), that is the GR components.
Actually the process can be improved
considering all the components that we have considered here. Before
starting with the analysis, it has to be emphasized that,
considering a stochastic background of  GWs, it can be
described and characterized by
a dimensionless spectrum (see the definition
\cite{Allen,AO,Maggiore,Grishchuk})

\begin{equation}
\Omega^A_{gw}(f)=\frac{1}{\rho_{c}}\frac{d\rho^A_{gw}}{d\ln
f},\label{eq: spettro}\end{equation} where
 \begin{equation}
\rho_{c}\equiv\frac{3H_{0}^{2}}{8\pi G}\label{eq: densita'
critica}\end{equation} is the (actual) critical energy density of
the Universe, $H_0$ the today observed Hubble expansion rate, and
$d\rho_{sgw}$ is the energy density of the part of the
gravitational radiation contained in the frequency range $f$ to
$f+df$.
\begin{equation}
\rho_{\rm gw}=\int_0^{\infty}df\,\tilde{\rho}_{\rm gw}(f)\,.
\end{equation}
 where $\tilde{\rho}_{\rm GW}$
is the GWs energy density per unit frequency.
 $\Omega_{\rm{gw}}(f)$ is related to $S_h(f)$ by \cite{Allen2,Maggiore}
\begin{equation}
\Omega_{\rm{gw}}^A (f) = \left( \frac{4 \pi^2}{3H_0^2} \right) f^3 S_h^A (f)\:.
\label{eq22}
\end{equation}
Note that the above definition is different from that in the
literature \cite{Allen2,Maggiore}, by a factor of 2, since it is
defined for each polarization. It is convenient to represent the
energy density with the form $h_{0}^2 \, \Omega_{\rm{gw}}(f)$ by
parametrizing the Hubble constant as $H_0=100\,h_{0}\,\rm{km\,
s^{-1}\,Mpc^{-1} }$. Then, the GW stochastic background energy density of all modes can
be written as
\begin{eqnarray}
\Omega_{\rm{gw}}^{A} \equiv  \Omega_{\rm{gw}}^{+} + \Omega_{\rm{gw}}^{\times}+
 \Omega_{\rm{gw}}^B + \Omega_{\rm{gw}}^C +  \Omega_{\rm{gw}}^D+\Omega_{\rm{gw}}^s\nonumber\\
 \label{eq23}
\end{eqnarray}
we can split $\Omega_{\rm{gw}}^{A}$ as a part arising from GR
\begin{equation}
\Omega_{\rm{gw}}^{GR}= \Omega_{\rm{gw}}^{+} + \Omega_{\rm{gw}}^{\times}\, , \qquad  \Omega_{\rm{gw}}^{+} = \Omega_{\rm{gw}}^{\times}\label{GR}
\end{equation}
a part from higher-order-gravity
\begin{equation}
\Omega_{\rm{gw}}^{HOG}= \Omega_{\rm{gw}}^B + \Omega_{\rm{gw}}^C +  \Omega_{\rm{gw}}^D\, , \qquad  \Omega_{\rm{gw}}^B = \Omega_{\rm{gw}}^C = \Omega_{\rm{gw}}^D\label{HOG}
\end{equation}
and a scalar part
$\Omega_{\rm{gw}}^s$.

We are considering now standard units and study only the modes
which arise from higher order theory.

The relic stochastic background of GWs can be derived by
considering only general assumptions and basic principles of
Quantum Field Theory and GR. The quantum fluctuations of the
zero-point energy can be amplified in the early Universe by the
large variations of gravity and this mechanism produces GWs. A
very interesting by-product of GWs is that they can be used to
probe the evolution of the Universe at early times, even up to the
Planck epoch and the Big Bang singularity
\cite{Allen,AO,Maggiore,Grishchuk}. The mechanism of the GWs is
connected to inflationary scenario \cite{Watson,Guth}, which fits
well the WMAP data and is in particularly good agreement with
almost exponential inflation and spectral index $\approx1$,
\cite{Bennet,Spergel}.

A remarkable fact about the inflationary scenario is that it
contains a natural mechanism which gives rise to perturbations for
any field. It is important for our aims that such a mechanism
provides also a distinctive spectrum for relic scalar GWs. These
perturbations in inflationary cosmology arise from the most basic
quantum mechanical effect: the uncertainty principle. In this way,
the spectrum of relic GWs that we could detect today is nothing
else but the adiabatically-amplified zero-point fluctuations
\cite{Allen,Grishchuk}. The calculation for a simple inflationary
model can be performed for the scalar field component  of eq.
(\ref{eq: perturbazione totale}). Let us assume that the early
Universe is described an inflationary de Sitter phase emerging in
a radiation dominated phase \cite{Allen,AO,Grishchuk}. The
conformal metric element is
\begin{equation}
ds^{2}=a^{2}(\eta)[-d\eta^{2}+d\overrightarrow{x}^{2}+h_{\mu\nu}(\eta,\overrightarrow{x})dx^{\mu}dx^{\nu}],\label{eq: metrica}\end{equation}
where, for a purely  GW the metric perturbation (\ref{eq: perturbazione totale})
reduces to
\begin{equation}
h_{\mu\nu}=h_{A} e_{\mu\nu}^{(A)}.\label{eq: perturbazione
scalare}\end{equation}
where  $A=+,\times,B,C,D,$ and $s$.
Let us assume a phase transition between
a de Sitter  and a radiation-dominated phase \cite{Allen,Grishchuk}, we have:
$\eta_1$ is the inflation-radiation transition conformal time and
$\eta_0$ is the value of conformal time today.  If we express the
scale factor in terms of comoving time $cdt=a(t)d\eta$,
 we have
\begin{equation}
\label{eq:dominioradiazione}
a(t)\propto\exp(H_{ds}t), \qquad a(t)\propto\sqrt{t}
\end{equation}
 for the de Sitter and radiation phases respectively. In order to solve
 the horizon and flatness problems, the condition
${\displaystyle \frac{a(\eta_{0})}{a(\eta_{1})}>10^{27}}$ has to
be satisfied. The relic scalar-tensor GWs are the weak
perturbations $h_{\mu\nu}(\eta,\overrightarrow{x})$ of the metric
(\ref{eq: perturbazione scalare}) which can be written in the form
\begin{equation}
h_{\mu\nu}=e_{\mu\nu}^{(A)}(\hat{k})X(\eta)\exp(i \overrightarrow{k}\cdot\overrightarrow{x}),\label{eq:
relic gravity-waves}\end{equation} in terms of the conformal time
$\eta$ where $\overrightarrow{k}$ is a constant wavevector. From
eq.(\ref{eq: relic gravity-waves}), the component is
\begin{equation}
\Phi(\eta,\overrightarrow{k},\overrightarrow{x})=X(\eta)\exp(i \overrightarrow{k}\cdot\overrightarrow{x}).\label{eq:
phi}\end{equation} Assuming $Y(\eta)=a(\eta)X(\eta)$, from the
Klein-Gordon equation in the FRW metric, one gets
\begin{equation}
Y''+\left(|\overrightarrow{k}|^{2}-\frac{a''}{a}\right)Y=0\label{eq:
Klein-Gordon}\end{equation} where the prime $'$ denotes derivative
with respect to the conformal time. The solutions of eq. (\ref{eq:
Klein-Gordon})  can be expressed in terms of Hankel functions in
both the
inflationary and radiation dominated eras, that is:\\
For $\eta<\eta_{1}$ \begin{equation}
X(\eta)=\frac{a(\eta_{1})}{a(\eta)}[1+i H_{ds}\omega^{-1}]\exp \left(-ik(\eta-\eta_{1}) \right),\label{eq:
ampiezza inflaz.}\end{equation}
for $\eta>\eta_{1}$
\begin{equation}
X(\eta)=\frac{a(\eta_{1})}{a(\eta)} \left[ \alpha\exp\left(-ik(\eta-\eta_{1}) \right)+\beta\exp
\left( ik(\eta-\eta_{1}) \right) \right],\label{eq: ampiezza rad.}\end{equation} where
$\omega=ck/a$ is the angular frequency of the wave (which is
function of the time being $k=|\overrightarrow{k}|$ constant),
$\alpha$ and $\beta$ are time-independent constants which we can
obtain demanding that both $X$ and $dX/d\eta$ are continuous at
the boundary $\eta=\eta_{1}$ between the inflationary and the
radiation dominated eras. By this constraint, we obtain
\begin{equation}
\alpha=1+i\frac{\sqrt{H_{ds}H_{0}}}{\omega}-\frac{H_{ds}H_{0}}{2\omega^{2}}\,,\qquad
\beta=\frac{H_{ds}H_{0}}{2\omega^{2}}\label{eq:
beta}\end{equation} In eqs. (\ref{eq: beta}),
$\omega=ck/a(\eta_{0})$ is the angular frequency as observed
today, $H_{0}=c/\eta_{0}$ is the Hubble expansion rate as observed
today. Such calculations  are referred in  literature as the
Bogoliubov coefficient methods \cite{Allen,Grishchuk}.

In an inflationary scenario, every  classical or macroscopic
perturbation is damped out by the  inflation, i.e. the minimum
allowed level of fluctuations is that required by the uncertainty
principle. The  solution (\ref{eq: ampiezza inflaz.}) corresponds
to a de Sitter vacuum state. If the period of inflation is long
enough, the today observable properties of the Universe  should be
indistinguishable from the properties of a Universe started in the
de Sitter vacuum state. During the radiation dominated phase, the
particles are described by the eigenmodes that correspond to the
coefficients of $\alpha$, while the antiparticles correspond to
the coefficients of $\beta$. Therefore, the number of  particles
that have been created at angular frequency $\omega$ in the
radiation phase is given by
\begin{equation}
N_{\omega}=|\beta_{\omega}|^{2}=\left(\frac{H_{ds}H_{0}}{2\omega^{2}}\right)^{2}.\label{eq:
numero quanti}\end{equation} Now it is possible to write an
expression for the energy density of the stochastic scalar-tensor
relic gravitons background in the frequency interval
$(\omega,\omega+d\omega)$ for each mode as
\begin{equation}
d\rho_{gw}^A=\hbar\omega\left(\frac{\omega^{2}d\omega}{2\pi^{2}c^{3}}\right)N_{\omega}=
\frac{\hbar
H_{ds}^{2}H_{0}^{2}}{8\pi^{2}c^{3}}\frac{d\omega}{\omega}=\frac{\hbar
H_{ds}^{2}H_{0}^{2}}{8\pi^{2}c^{3}} \frac{df}{f}\,,\label{eq: de
energia}\end{equation} where $f$, as above, is the frequency in
standard comoving time. eq. (\ref{eq: de energia}) can be
rewritten in terms of the today and de Sitter value of energy
density being
\begin{equation} H_{0}^2=\frac{8\pi G\rho_{c}}{3c^{2}}\,,\qquad H_{ds}^2=\frac{8\pi G\rho_{ds}}{3c^{2}}.\end{equation}
Introducing the Planck density ${\displaystyle
\rho_{Planck}=\frac{c^{7}}{\hbar G^{2}}}$ the spectrum is given by
\begin{equation}
\Omega_{gw}^{A}(f)=\frac{1}{\rho_{c}}\frac{d\rho_{gw}}{d\ln
f}=\frac{f}{\rho_{c}}\frac{d\rho_{gw}}{df}=\frac{8}{9}\frac{\rho_{ds}}{\rho_{Planck}}.\label{eq:
spettro gravitoni}\end{equation} At this point,  some  comments
are in order. First of all, such a calculation works for a
simplified model that does not include the matter dominated era.
If we also include such an era, we would also have to take into
account the redshift at the equivalence epoch and this results in
\cite{Allen2}
\begin{equation}
\Omega_{gw}^{A}(f)=\frac{8}{9}\frac{\rho_{ds}}{\rho_{Planck}}(1+z_{eq})^{-1},\label{eq:
spettro gravitoni redshiftato}\end{equation} for the waves which,
at the epoch in which the Universe becomes matter dominated, have
a frequency higher than $H_{eq}$, the Hubble parameter at
equivalence. This situation corresponds to frequencies
$f>(1+z_{eq})^{1/2}H_{0}$. The redshift correction in eq.(\ref{eq:
spettro gravitoni redshiftato}) is needed since the today observed
Hubble parameter $H_{0}$ would result different  without a matter
dominated contribution. At lower frequencies, the spectrum is
given by \cite{Allen,Grishchuk}
\begin{equation}
\Omega_{gw}(f)\propto f^{-2}.\label{eq: spettro basse
frequenze}\end{equation} As a further consideration, let us note
that the results (\ref{eq: spettro gravitoni}) and (\ref{eq:
spettro gravitoni redshiftato}), which are not frequency
dependent, do not work correctly in all the range of physical
frequencies. Waves that have frequencies less than $H_{0}$, the
energy density is in a sense not well defined, as their wavelength
becomes larger than the Hubble scale of the Universe. In a similar
manner, at high frequencies, there is a maximal frequency above
which the spectrum rapidly drops to zero. In the above
calculation, the simple assumption that the phase transition from
the inflationary to the radiation dominated epoch is instantaneous
has been made. In the physical Universe, this process occurs over
some time scale $\Delta\tau$, being
\begin{equation}
f_{max}=\frac{a(t_{1})}{a(t_{0})}\frac{1}{\Delta\tau},\label{eq:
freq. max}\end{equation} which is the redshifted rate of the
transition. In any case, $\Omega_{gw}^{A}$ drops rapidly. The two
cutoffs at low and high frequencies for the spectrum guarantee
that the total energy density of the relic  gravitons is finite.
These results can be quantitatively constrained considering the
recent WMAP release. Nevertheless, since the spectrum falls off
$\propto f^{-2}$ at low frequencies, this means that today, at
LIGO-VIRGO and LISA frequencies, one gets for the GR part
\cite{Maggiore,Buonanno:2003th}
\begin{equation} \Omega_{gw}^{GR}(f)h_{100}^{2}<2 \times
10^{-6}.\label{eq: limite spettroGR}\end{equation}
for the higher-order-gravity part
\begin{equation} \Omega_{gw}^{HOG}(f)h_{100}^{2}<6.7\times
10^{-9}.\label{eq: limite spettroHOG}\end{equation}
and for the scalar part
\begin{equation} \Omega_{gw}^{s}(f)h_{100}^{2}<2.3\times
10^{-12}.\label{eq: limite spettro WMAP}\end{equation}

It is interesting to calculate the  corresponding strain at
$\approx 100Hz$, where interferometers like VIRGO and LIGO reach a
maximum in sensitivity \cite{ligo,virgo}. With a minor
modification we can use the well known equation for the
characteristic amplitude \cite{Maggiore} for one of the components
of the GWs \footnote{The difference between our result and
eq.~(19) in Ref.~\cite{Maggiore} is due to the fact that the
latter did their calculation assuming the two polarization modes
of GR while we handle each mode separately, hence the
$\frac{1}{\sqrt{2}}$ difference.}:
\begin{equation}
h_{A}(f)\simeq8.93\times10^{-19}\left(\frac{1Hz}{f}\right)\sqrt{h_{100}^{2}\Omega_{gw}(f)},\label{eq:
legame ampiezza-spettro}\end{equation} and then we obtain for the
GR modes
\begin{equation}
h_{GR}(100Hz)<1.3\times 10^{-23}.\label{eq: limite per lo
strain}\end{equation}
while for the higher-order modes
\begin{equation}
h_{HOG}(100Hz)<7.3\times 10^{-25}.\label{eq: limite per lo
strain002}\end{equation} and for scalar modes
\begin{equation}
h_{s}(100Hz)<2\times 1.4 10^{-26}.\label{eq: limite per lo
strain003}\end{equation}

Then, since we expect a sensitivity of the order of $10^{-22}$ for
the above interferometers at $\approx100Hz$, we need to gain at
least three orders of magnitude. At smaller frequencies the
sensitivity of the VIRGO interferometer is of the order of
$10^{-21}$ at $\approx10Hz$ and in that case it is for the GR
modes
\begin{equation}
h_{GR}(100Hz)<1.3 \times 10^{-22}.\label{eq: limite per lo
strain01}\end{equation} while for the higher-order modes
\begin{equation}
h_{HOG}(100Hz)<7.3 \times 10^{-24}.\label{eq: limite per lo
strain02}\end{equation} and for scalar modes
\begin{equation}
h_{s}(100Hz)<1.4 \times 10^{-25}.\label{eq: limite per lo
strain03}\end{equation} \\
Still, these effects are below the sensitivity threshold to be
observed. The sensitivity of the LISA interferometer will be of
the order of $10^{-22}$ at $\approx 10^{-3} Hz$ (see \cite{lisa})
and in that case it is
\begin{equation}
h_{GR}(100Hz)<1.3 \times 10^{-18}.\label{eq: limite per lo
strain1}\end{equation} while for the higher-order modes
\begin{equation}
h_{HOG}(100Hz)<7.3 \times 10^{-20}.\label{eq: limite per lo
strain2}\end{equation} and for scalar modes
\begin{equation}
h_{s}(100Hz)<1.4 \times 10^{-21}.\label{eq: limite per lo
strain3}\end{equation} \\
This means that a stochastic background of
relic GWs could be, in principle, detected by the LISA
interferometer, including the additional modes.
\section{Conclusions}
\label{sei}

Our analysis covers extended  gravity models with a generic class
of Lagrangian density with higher order and terms of the form
$f(R,P,Q)$, where $ P\equiv R_{ab}R^{ab}$ and $Q\equiv
R_{abcd}R^{abcd}$. We have linearized the field equations for this
class of theories around a Minkowski background and found that,
besides a massless spin-2 field (the graviton), the theory
contains also spin-0 and spin-2 massive modes with the latter
being, in general,  ghosts. Then, we have investigated the
detectability of additional polarization modes of a stochastic GW
with ground-based laser-interferometric detectors and
space-interferometers. Such polarization modes, in general, appear
in the extended theories of gravitation and can be utilized to
constrain the theories beyond GR in a model-independent way.

However, a point has to be discussed in detail. If the
interferometer is directionally sensitive and we also know the
orientation of the source (and of course if the source is
coherent) the situation is straightforward. In this case, the
massive mode coming from the simplest extension, $f(R)$-gravity,
would induce longitudinal displacements along the direction of
propagation which should be detectable and only the amplitude due
to the scalar mode would be the true, detectable, "new" signal
\cite{felix}. But even in this case, we could have  a second
scalar mode inducing a similar effect, coming from the massive
ghost, although with a minus sign. So in this case, one  has
deviations from the prediction of  $f(R)$-gravity, even if  only
the massive modes are considered as new signal.

On the other hand, in the case of the stochastic background, there
is no coherent source and no directional detection of the
gravitational radiation. What the interferometer picks is just an
averaged signal coming from the contributions of all possible
modes from (uncorrelated) sources all over the celestial sphere.
Since we expect the background to be isotropic, the signal will be
the same regardless of the orientation of the interferometer, no
matter how or on which plane it is rotated, it would always record
the characteristic amplitude $h_c$. So there is intrinsically no
way to disentangle any of the mode in the background, being $h_c$
related to the total energy density of the gravitational
radiation, which  depends on the number of modes available. Every
mode, essentially, contributes in the same manner, at least in the
limit where the mass for the massive and ghost modes are very
small (as they should be). So, it should be the number of the
modes available that makes the difference, not their origin.

Again, even if this does not hold, one should still get into
consideration at least the massive ghost mode to get a constraint.
This is the why we have considered only $h_{GR}$, $h_{HOG}$ and
$h_{s}$ in the above cross-correlation analysis without giving
further fine details coming from polarization. For the situation
considered here, we find that the massive modes are certainly of
interest for direct attempts at detection with the LISA
experiment. It is, in principle, possible that massive GW modes
could be produced in more significant quantities in cosmological
or early astrophysical processes in alternative theories of
gravity,  being this possibility still unexplored. This situation
should be kept in mind when looking for a signature distinguishing
these theories from GR, and seems to deserve further
investigation.

\section*{Acknowledgements}
We would like to thank H. Collins, R. Boels,  C. Charmousis and L.
Milano for useful discussions and comments. C.B. is supported by
the CNRS and the Universit\'e de Paris-Sud XI. M.D. acknowledges
the support by VIRGO collaboration.  S.N. acknowledges the support
by the Niels Bohr International Academy, the EU FP6 Marie Curie
Research $\&$ Training Network ``UniverseNet" under Contract No.
MRTN-CT-2006-035863 and the Danish Research Council under FNU
Grant No. 272-08-0285.


\end{document}